\documentclass[fleqn,11pt]{article}
\usepackage{graphicx}% Include figure files
\usepackage{amsfonts}

\newcommand{\bq}{\begin{equation}}            
\newcommand{\eq}{\end{equation}}            
\newcommand{\bqa}{\begin{eqnarray}}            
\newcommand{\eqa}{\end{eqnarray}}            
\newcommand{\ba}{\begin{array}} 
\newcommand{\ea}{\end{array}} 
\newcommand{\bi}{\begin{itemize}} 
\newcommand{\ei}{\end{itemize}} 

\newcommand{\al}{\alpha}                
\newcommand{\be}{\beta}                
\newcommand{\ga}{\gamma}               
\newcommand{\dt}{\delta}

\newcommand{\si}{\sigma}

\newcommand{\ep}{\epsilon}

\newcommand{\cf}{{\cal F}}
\newcommand{\ck}{{\cal K}}

\newcommand{\nb}{{\nabla}}
\newcommand{\vs}{\vspace{2mm}}
\newcommand{\hs}{\hspace{5mm}}
\newcommand{\ra}{\rightarrow}           
\newcommand{\Pro}{\mbox{Prob}}
\newcommand{\df}{\stackrel{\triangle}{=}}     

\begin{document}
%\renewcommand{\thepage}{\roman{page}}
%\pagenumbering{roman}
%\baselineskip=0.65cm
%\baselineskip=0.6cm
\topmargin=-0.5in

\begin{center}
{\Large \bf A New Parameterized Family of Stochastic Particle Flow Filters}\\[2mm]
\end{center}
\begin{center}
\begin{tabular}{c  c}
        Liyi Dai & Fred Daum\\
        Raytheon Missiles \& Defense & \hspace{10mm} Raytheon Missiles \& Defense \\
        50 Apple Hill Drive & 235 Presidential Way \\
        Tewksbury, MA 01876 & Woburn, MA 01801 \\
        liyi.dai@raytheon.com & daum@raytheon.com 
\end{tabular}
\end{center}
\vs

\begin{abstract}
In this paper, we are interested in obtaining answers to the following questions for particle flow filters: Can we provide a theoretical guarantee that particle flow filters give correct results such as unbiased estimates? Are particle flows stable and under what conditions? Can we have one particle flow filter, rather than multiple seemingly different ones?  To answer these questions, we first derive a parameterized family of stochastic particle flow filters, in which particle flows are driven by a linear combination of prior knowledge and measurement likelihood information. We then show that several particle flows existing in the literature are special cases of this family. We prove that the particle flows are unbiased under the assumption of linear measurement and Gaussian distributions, and that estimates constructed from the stochastic flows are consistent. We further establish several finite time stability concepts for this new family of stochastic particle flows. The results reported in this paper represent a significant development toward establishing a theoretical foundation for particle flow filters.
\end{abstract}


\vs%pace{1cm}

%\centerline{\today}
\centerline{March 16, 2021, revised September 27, 2021}

%\newpage

\section{Introduction}
A number of general purpose filters are available for Bayesian filtering in the literature. The parameterized Bayesian filters include the Extended Kalman Filters \cite{Kal}, the Cubature Kalman Filters \cite{AH}, the Unscented Kalman Filters \cite{JU}, the  Gauss–Hermite filters \cite{DSG}, the central difference filters \cite{NOR}, and the Gaussian sum filters \cite{AS,IX}. The non-parameterized Bayesian filters include the particle filters \cite{AMGC}, the ensemble Kalman filters \cite{EVE}, the particle flow filters \cite{DH2007}, and other exact nonlinear filters \cite{Daum,LJ}. A number of comprehensive surveys or tutorials of these different methods are available in the literature, e.g., \cite{AMGC,CR,RAG, SLB}, and are beyond the scope of this paper. This paper focuses on a specific type of non-parametric density based nonlinear filters-the particle flow filters.

Particle filters have been widely used to solve a wide range of nonlinear filtering problems. Particle filters are sequential Monte Carlo methods that approximate conditional distributions by weights associated with a set of random sample points. The weights are updated sequentially. A long-standing problem with particle filtering is “particle degeneracy", referring to a situation in which all but usually one particle weights are close to zero. There have been numerous attempts to address this problem such as resampling. However, such efforts have not adequately resolved the problem of particle degeneracy \cite{DJ,PS,RAG}. For example, resampling introduces a new problem of particle impoverishment: During resampling particles with large weights are likely to be drawn multiple times whereas particles with small weights are not likely to be drawn at all, leading to a lack of particle diversity. Moreover, particle filters generally suffer from the “curse of dimensionality" as discussed in \cite{DH2011,QMG}.

The particle flow filtering was first introduced in \cite{DH2007} to mitigate the problem of particle degeneracy in particle filters. Instead of updating weights at fixed sample points of states (particles), particle flow filters seek to move all particles along dynamic flows (which will be described in more detail later), which effectively avoids the issue of particle degeneracy. It has been reported that filtering performance of particle flow filters is among the top ones in a wide range of applications \cite{CP,DOT,GYM,KKS,KB,MC,MCC,PC,PMZC,Port,SLC,Wu,Zhao}. 
%Benefits of the particle flow filters include (1) no resampling of particles is needed, thereby avoiding a major computational burden of particle filters and also the problem of particle impoverishment; (2) no exact forms of distributions are needed, which increases numerical accuracy and efficiency; (3) explicitly computing the conditional probability density is not needed, which increases numerical stability; (4) no weight updating is performed, thus avoids the problem of “particle degeneracy”; (5) no update of distributions is needed as in particle filters, which brings numerical efficiency and stability \cite{DHN2010}. Implementations of particle flow filters to practical tracking problems have indeed confirmed the realization of those benefits. Particle flow filters offer orders of magnitude improvements in accuracy and/or speed over conventional Extended Kalman Filters and particle filters while avoiding "particle degeneracy" and "particle impoverishment" \cite{DHN2010}.
Despite abundant empirical evidences showing the top performance, a theoretical foundation for particle flow filters is yet to be developed. %not to mention at a level comparable with those for the Extended Kalman Filters or particle filters. 
Part of the reason is that particle flow filters are motivated to address issues in nonlinear filtering which has proven to be highly challenging for theoretical analysis. The relatively short history of particle flow filters may be another contributing factor.

This paper is intended to serve two purposes. First, a number of particle flows have been proposed in the literature for the implementation of particle flow filters \cite{DH2015}. Those flows offer different benefits, but appear disconnected from each other. In this paper, we derive a new general form of particle flows as a parameterized family of stochastic flows. The “parameter" is resulted from introducing a nonzero diffusion process to drive the flows as a scheme to improve numerical stability in implementation. We show that the new form of particle flows includes all linear flows driven by a non-zero diffusion process with noises independent of the state. Several existing stochastic flows are special cases of this general form, thus providing a unifying form of stochastic particle flows.

Our second motivation is the desire to build a theoretical foundation for the particle flow filtering framework. Significant amount of theoretical studies has been devoted to Kalman filters and particle filters over the last sixty years since R.E. Kalman proposed the Kalman filter in \cite{Kal}. A theoretical foundation is yet to be established for particle flow filters. We start with basic properties in estimation and filtering: unbiasedness, consistency and numerical stability. We show that the particle flows are unbiased under the assumption of Gaussian distributions and that estimates constructed from particle flows are consistent. Connection with the broadly-adopted linear minimum variance estimation is established. Our contributions are significant because they provide a theoretical guarantee that particle flow filters indeed give correct answers. For numerical stability, we adopt a Lyapunov-like approach to the stability analysis of particle flows. Particularly, we establish bounded error stability for general stochastic flows, and provide mild conditions that ensure contractive stability. Conventional concepts of stability are concerned with asymptotic behaviors as time goes to infinity. Particle flows are derived from homotopy with finite “time". Finite time stability concepts are adopted in this paper while we also provide quantitative characterization on the dynamic behaviors of error system.

%Theoretical analysis of nonlinear filters has been a challenging task. Traditionally, for both Kalman filters and particle filters, the strategy is to carry out theoretical analysis under the assumptions of Gaussian distributions and linear measurements as a local approximation to nonlinear problems. As an initial effort, we adopt the same strategy whenever applicable and needed.

Particle flow filters are constructed for sequential state estimation, updating state estimate as new data is collected. To keep notations as simple as possible and without loss of generality, we are mainly concerned with one-step Bayesian estimation. In practice, its implementation is sequential consisting of multiple one-step estimation. The analysis is applicable to general Bayesian inference problems that go beyond filtering.

The rest of the paper is organized as the following. In Section 2, we derive a parameterized family of stochastic particle flow filters. In Sections 3, we establish basic statistical properties of the stochastic particle flows. In Section 4, we show that several stochastic flows and one deterministic flow existing in the literature can be recast as special cases of the parameterized family of stochastic flows we derive. In Section 5, we establish several Lyapunov-like stability results for error system in numerical implementation. To keep continuity of reading, all proofs are moved to the Appendix.
\vs

\noindent
{\em Notations}.

We use $\mathbb{R}^n$ to denote the real valued $n$ dimensional Euclidean space, $\mathbb{R}=\mathbb{R}^1$, $\mathbb{R}^+$ is the set of non-negative real numbers, and $\mathbb{R}^{n\times m}$ is the real valued $n\times m$ matrix space.

We use lowercase letters to denote column vectors or scalars, uppercase letters to denote matrices. An identity matrix is $I$ or $I_{n\times n}$ if we need to specify its $n\times n$ dimension. For a vector $a$ or matrix $A$, its transpose is $a^T$ or $A^T$, respectively. For symmetric matrices $A$ and $B$, $A\geq 0$ or $A>0$ denotes that $A$ is positive semi-definite or positive definite, respectively, and $A\geq B$ or $A>B$ denotes $A-B\geq 0$ or $A-B>0$, respectively. For a square matrix $A$, its determinant is $|A|$, its minimum eigenvalue is $\lambda_{min}(A)$. 

For a random variable $x\in \mathbb{R}^n$, its mean is $E[x]$. For a scalar function $f(x): \mathbb{R}^n\ra \mathbb{R}$, its gradient is $\nb_x f(x)=[\partial f/\partial x_1, \partial f/\partial x_2, ..., \partial f/\partial x_n]^T\in\mathbb{R}^n$, and its divergence is $div(f) = \sum_{i=1}^n\partial f/\partial x_i\in\mathbb{R}$.

Finally, we use “$A\Rightarrow B$" as a concise form of the statement “$A$ leads to $B$".

\section{Derivation of Parameterized Stochastic Particle Flows}
Assume a given probability space, on which a random variable $x\in \mathbb{R}^n$ is defined and takes values in a $n$ dimensional real space $\mathbb{R}^n$ and a measurement of $x$, $z\in \mathbb{R}^d$, defined in a $d$ dimensional real space. Let $p_x(x)$ denote the prior probability density function of $x$ and $p_z(z|x)$ the likelihood of a measurement $z$. The Bayes' Theorem states that the posterior conditional density function of $x$ for a given measurement $z$, $p_x(x|z)$, is given by\footnote{To keep notations as simple as possible, in this paper we focus on one-step Bayesian estimation which can be applied to filtering or inference problems. For multi-step sequential filtering, the Bayes' Theorem is as the following \cite{DH2007,Jaz}.
\[
p(x,t_k|Z_k)=p(z_k|x,t_k)p(x,t_k|Z_{k-1})/p(z_k|Z_{k-1})
\]
in which $z_k$ is the $k$-th measurement at time $t_k$, $Z_k=\{z_1,z_2,...,z_k\}$, and $p(z_k|x,t_k)$ is the probability density of measurement $z_k$ at time $t_k$ conditioned on $x$. The probability density functions $g(x)$, $h(x)$, and $p(x)$ in (\ref{homotopy}) need to be replaced, respectively, with the following.
\[ 
p(x)=p(x,t_k|Z_k), \hs g(x)=p(x,t_k|Z_{k-1}), \hs h(x)=p(z_k|x,t_k).
\]
The rest of discussion follows.
}
\bq
p_x(x|z) = \frac{p_x(x)p_z(z|x)}{p_z(z)}
\label{bayes}
\eq
in which $p_z(z)=\int_x p_z(z|x)p_x(x)dx$ is also known as the normalization factor. Without loss of generality, it is assumed throughout this paper that all probability density functions exist, sufficiently (second order) differentiable, and are non-vanishing everywhere. Otherwise we restrict discussions to the supports of the density functions.

For simplicity, we denote
\[ p(x)=p_x(x|z), \hs g(x) = p_x(x), \hs h(x) = p_z(z|x).
\]
In the framework of particle flow filters, particle flows are defined through homotopy. Toward that end, we define a new conditional probability density function as the following 
\bq
p(x,\lambda) = \frac{g(x)h^\lambda(x)}{c(\lambda)}
\label{homotopy}
\eq
for all $\lambda\in[0, 1]$. In (\ref{homotopy}), $c(\lambda)$ is the normalization factor so that $p(x,\lambda)$ remains a probability density function for all $\lambda\in[0, 1]$. It's clear from (\ref{homotopy}) that 
\[    p(x,0) = g(x), \hs p(x,1)=p(x). \]
In other words, $p(x,0)$ is the density function of the prior distribution and $p(x,1)$ is that of the posterior distribution. Therefore, the mapping $p(x,\lambda): \mathbb{R}^+\times [0, 1]\longrightarrow \mathbb{R}^+$ in (\ref{homotopy}) defines a homotopy from $g(x)$ to $p(x)$. By taking the natural logarithm on both sides of (\ref{homotopy}), we obtain
\bq
\log p(x,\lambda) = \log g(x)+ \lambda \log h(x) -\log c(\lambda).
\label{loghom}
\eq
Recall that a major problem with particle filters is particle degeneracy \cite{DH2007,DJ,PS,RAG}. To mitigate this issue, particle flow filters move (change) $x$ as a function of $\lambda$, $x(\lambda)$, so that (\ref{homotopy}), or equivalently (\ref{loghom}), is always satisfied as $\lambda$ changes from $0$ to $1$. The value of $x(\lambda)$ at $\lambda=1$ is used for estimation in problems such as filtering or Bayesian inference. It turns out that there exists much freedom in the choice of $\{x(\lambda), \lambda\in[0, 1]\}$ \cite{DH2015}.
The $x(\lambda)$ could be driven by a deterministic process as in the Exact Flow \cite{DHN2010,DHN2018}, or by a stochastic process as in stochastic flows \cite{DH2013,DHN2016,DHN2018}. In this paper, we consider a stochastic flow in which $x(\lambda)$ is driven by the following stochastic process
\bq
dx = f(x,\lambda)d\lambda+q(x,\lambda)dw_{\lambda}
\label{flow}
\eq
where $f(x,\lambda)\in \mathbb{R}^n$ is a drift function, $q(x,\lambda)\in \mathbb{R}^{n\times m}$ is a diffusion matrix, and $w_{\lambda}\in \mathbb{R}^m$ is a $m$ dimensional Brownian motion process in $\lambda$ with $E[dw_\lambda dw_\lambda^T ]=\sigma(\lambda)d\lambda$. The stochastic differential equation (\ref{flow}) is a standard diffusion process \cite{Jaz}. Note that $\{x(\lambda), \lambda \in [0, 1]\}$ is a stochastic process in $\lambda$. In this problem formulation, the diffusion matrix $q(x,\lambda)$ serves as a design parameter which should not be confused with the process noise matrix of the underlying stochastic system. For clarity, we drop its dependence on $\lambda$ but add the dependence back when it is beneficial or necessary to emphasize its dependence on $\lambda$. Without loss of generality, we assume that $\sigma(\lambda)=I_{m\times m}$. We denote
\[ Q(x,\lambda) = q(x,\lambda)q(x,\lambda)^T \in \mathbb{R}^{n\times n}. \]
Note that the matrix $Q(x,\lambda)$ is always symmetric positive semi-definite for any $x$ and $\lambda$. Again, this matrix $Q$ should not be confused with the covariance matrix of the process noise of the underlying system.

Our goal is to select $f(x,\lambda)$ and $q(x,\lambda)$, or equivalently $Q(x,\lambda)$, such that (\ref{loghom}) is maintained  for the particle $x(\lambda)$ driven by the stochastic process (\ref{flow}) for all $\lambda\in [0, 1]$. To that end, we start with the following lemma.
\vs

{\sc Lemma 2.1.}\cite{DH2013} For the particle flow $x(\lambda)$ defined in (\ref{flow}) to satisfy (\ref{loghom}), $f(x,\lambda)$ and $Q(x,\lambda)$ must satisfy the following condition
\bq
\nb_x\log h =-(\nb_x\nb_x^T\log p)f - \nb_x div(f)-(\nb_x f)^T(\nb_x\log p)+\nb_x[\frac{1}{2p}\nb_x^T(pQ)\nb_x] 
\label{cond1}
\eq 
in which all derivatives are assumed to exist. For simplicity and without causing confusion, in (\ref{cond1}) and for the rest of discussion in this paper, we omit all variables involved.
\vs

Since the introduction of particle flows in \cite{DH2007}, there have been steady efforts in the literature either to solve (\ref{cond1}) for a special $Q$ or to find an approximate solution for general $Q$ \cite{Daum2016}. In this paper, we focus on finding the exact solution  $f$ for arbitrary symmetric positive semi-definite (or positive definite) matrix $Q$ as long as $Q$ is not a function of $x$. The matrix $Q$ could be a function of $\lambda$.
\vs

{\sc Theorem 2.1.} {\em Assume that
\bi
\item[(A1).] $\nb_x \log g$ and $\nb_x \log h$ are linear in $x$, and
\item[(A2).] $\nb_x \nb_x^T \log p$ is non-singular for all $\lambda \in [0, 1]$.
\ei
Then for any matrix $K(\lambda) \in \mathbb{R}^{n\times n}$, independent of $x$, (\ref{cond1}) is satisfied by the following $f$ and $Q$
\bq
f = (\nb_x\nb_x^T\log p)^{-1}[-\nb_x\log h+K(\nb_x\nb_x^T\log p)^{-1}(\nb_x\log p)],
\label{f1}
\eq
\bq
Q = (\nb_x\nb_x^T\log p)^{-1}(-\nb_x\nb_x^T\log h+K+K^T)(\nb_x\nb_x^T\log p)^{-1}.
\label{q1}
\eq
provided that $Q$ is positive semi-definite.}
\vs

Note that the $Q$ defined in (\ref{q1}) is always symmetric for any $K$ since $\nb_x\nb_x^T\log p$ and $\nb_x\nb_x^T\log h$ are symmetric by definition (recall that $g(x)$, $h(x)$, and consequently $p(x)$ are assumed sufficiently differentiable). 

Under the assumptions (A1) and (A2) in Theorem 2.1, (\ref{f1}) shows that the function $f$ is a linear combination of prior knowledge $\nb_x\log g$ and measurement likelihood information $\nb_x\log h$. The assumptions (A1) is satisfied if $g$ and $h$ are Gaussian or exponential.
% and the measurement equation is
%\bq
%dz = Hxd\lambda + dv
%\eq
%in which $H\in R^{d\times n}$ does not depend on $x$, and $dv$ is a diffusion process in $\lambda$. 
Such Gaussian assumption is widely adopted initially as a local approximation to nonlinear problems in the analysis of extended Kalman filters or particle filters, as well as in many fields of studies. It should be pointed out that (\ref{flow}), (\ref{f1}) and (\ref{q1}) define a general purpose nonlinear filter without the assumption (A1).% which only serves as a stepping-stone in the derivation.

Theorem 2.1 gives a family of stochastic particle flows $x(\lambda)$ parameterized by the matrix $K$. The following corollary states that for any matrix $Q(\lambda)\in R^{n\times n}$ independent of $x$, we can find a corresponding stochastic particle flow $\{x(\lambda), \lambda \in [0, 1]\}$.
\vs

{\sc Corollary 2.1.} {\em Assume the assumptions (A1) and (A2) in Theorem 2.1. For any symmetric positive semi-definite matrix $Q(\lambda)\in R^{n\times n}$, (\ref{cond1}) is satisfied by the $f$ defined in (\ref{f1}) with $K(\lambda)$ chosen as 
\bq
K(\lambda) = \frac{1}{2}(\nb_x\nb_x^T\log p)Q(\lambda)(\nb_x\nb_x^T\log p)+\frac{1}{2}(\nb_x\nb_x^T\log h).
\label{q2}
\eq
}
\vs

Corollary 2.1 states that we can find a drift function $f$ for any given matrix $Q$ as long as $Q$ is positive semi-definite and independent of $x$. In other words, $Q$ is a parameter matrix. It has been observed that a nonzero $Q$ could improve numerical stability for solving (\ref{flow}) in practice \cite{DH2014}. We will discuss in Section 5 how choices of $Q$ affect numerical stability of the particle flow (\ref{flow}) for practical implementation. In this sense, $Q$ acts as a stabilizer. 

Note that the drift function $f$ (\ref{f1}) is a linear combination of prior and measurement information for any $Q$. We may ask if there are other linear functions that may also solve (\ref{cond1}). Note that (\ref{f1}) is a parameterized solution. We could not simply adopt conventional definition of solution uniqueness which often refers to {\em the} solution. We need to define a new concept of solution equivalence.
%The following Theorem 2.2 gives a negative answer.
\vs

{\sc Definition 2.1.} {\em For given density functions $g(x)$ and $h(x)$, we define the constrained parameter space as $\ck\df\{K\in \mathbb{R}^{n\times n} | K+K^T-\nb_x\nb_x^T\log h \geq 0\}$, and the solution space $\cf$ as
\[
\cf\df\{f\in \mathbb{R}^n | f=(\nb_x\nb_x^T\log p)^{-1}[-\nb_x\log h+K(\nb_x\nb_x^T\log p)^{-1}(\nb_x\log p)], K\in \ck \}.
\]
Then $\cf$ actually defines a solution manifold.
}
\vs

{\sc Theorem 2.2.} {\em Assume the assumptions (A1) and (A2) in Theorem 2.1. Then for any linear solution to (\ref{cond1}) of the form $f=    A(\lambda)x+b(\lambda)$, we must have $f\in\cf$.}
\vs

Therefore, the drift function $f$ given by (\ref{f1}) and (\ref{q1}) is the "unique" linear solution to (\ref{cond1}) with possible difference in the choice of the parameter matrix $K$.

\section{Properties of Particle Flows}

When particle flows are used for example in filtering or Bayesian inference, one typically considers a number of particles, $\{x_i(\lambda), i=1, 2, ..., N\}$, starting with initial conditions (prior) and driven by the stochastic process (\ref{flow}). Then the average
\bq \hat{x}_N=\frac{1}{N}\sum_{i=1}^{N}x_i(\lambda)|_{\lambda=1} 
\label{est_mean}
\eq
could be used as an estimate of the posterior mean. The estimate of the posterior covariance matrix could be constructed as 
%the maximum likelihood estimate for Gaussian distributions 
\cite{MKB},
\bq
\hat{P}_N = \frac{1}{N-1}\sum_{i=1}^N (x_i(\lambda)-\hat{x}_N)(x_i(\lambda)-\hat{x}_N)^T|_{\lambda=1}.
\label{est_var}
\eq
%or in general,
%\[ \hat{P}_N = \frac{1}{N-1}\sum_{i=1}^N (x_i(\lambda)-\hat{x}_N)(x_i(\lambda)-\hat{x}_N)^T|_{\lambda=1}. \]
One question is whether the estimates are unbiased and their limiting behavior as the number of particles increases. 

Under the assumption (A1) in Theorem 2.1, $\nb_x\log g$ and $\nb_x\log h$ are linear in $x$. Therefore, we may re-write (\ref{flow}) as the following.
\bq
dx = [A(\lambda)x+b(\lambda)]d\lambda+q(x,\lambda)dw_{\lambda}
\label{lin1}
\eq
in which
\bq A(\lambda) = (\nb_x\nb_x^T\log p)^{-1}[-\nb_x\nb_x^T \log h+K],
\label{lin1_A}
\eq
\bq b(\lambda) = f-A(\lambda)x,
\label{lin1_b}
\eq
and
\[ Q(\lambda)= (\nb_x\nb_x^T\log p)^{-1}(-\nb_x\nb_x^T\log h+K+K^T)(\nb_x\nb_x^T\log p)^{-1}.
\]

The linear system (\ref{lin1}) defines a $\lambda$-varying linear stochastic differential equation from which we can derive its mean and covariance matrix for $\lambda\in [0, 1]$.
\vs

{\sc Lemma 3.1.}\cite{Arn,Jaz} {\em Let $\bar{x}(\lambda)=E[x]$ and $P(\lambda)=E[(x-\bar{x})(x-\bar{x})^T]$ be the mean and the covariance matrix of $x(\lambda)$, respectively. Then we have
\bq
\frac{d\bar{x}}{d\lambda} = A(\lambda)\bar{x}+b(\lambda), \hs \bar{x}|_{\lambda=0}=x_0
\label{mean1}
\eq
and
\bq
\frac{dP}{d\lambda} = A(\lambda)P+PA^T(\lambda)+Q(\lambda), \hs P|_{\lambda=0}=P_0.
\label{cov1}
\eq
where $x_0$ and $P_0$ are the mean and the covariance matrix of prior and determined by prior density function $g(x)$.}
\vs

The Gaussian distribution plays a critical role in the development of the theory of filtering and estimation. It has been broadly used in the modeling and analysis in virtually every field, either engineering, natural science, or social studies. %Difficult problems may become traceable under Gaussian assumptions. Filtering and estimation algorithms developed under Gaussian assumptions often work remarkably well in practice. 
We next establish that, under Gaussian assumptions, the particle flow defined in (\ref{f1})-(\ref{q1}) indeed has $p(x,\lambda)$ as its density function with correct mean and covariance matrix for all $\lambda\in[0, 1]$.
\vs

For a flow $x$ determined by the stochastic differential equation (\ref{flow}), its density $p(x,\lambda)$ satisfies the following Kolmogorov's forward equation (also known as the Fokker–Planck equation) \cite{Jaz}
\bq  
\frac{\partial p}{\partial\lambda}=-div(pf)+\frac{1}{2}\nb_x^T(pQ)\nb_x.
\label{kfe}
\eq
%In general, finding the analytical solution to a high dimensional Kolmogorov's forward equation is feasible for only special cases \cite{Ris}. 
If $g(x)$ and $h(x)$ are Gaussian distributed, we know that $p(x,\lambda)$ is also Gaussian according to (\ref{homotopy}). We only need to verify that it satisfies (\ref{kfe}). 
\vs

{\sc Theorem 3.1}. {\em Assume that the density function for the prior is Gaussian as the following.
\bq
g(x) = \frac{1}{\sqrt{(2\pi)^n|P_g|}} exp\{-\frac{1}{2} (x-x_{prior})^TP_g^{-1}(x-x_{prior})\}
\label{Gau_g}
\eq
and the measurement is linear
\bq
z = Hx+v
\label{lin_meas}
\eq
with $H\in \mathbb{R}^{d\times n}$ (independent of $x$), $v\in \mathbb{R}^d$ is Gaussian with zero mean and covariance matrix $E[vv^T]=R\in \mathbb{R}^{d\times d}$ that is positive definite.
%\bq
%h(x) = \frac{1}{\sqrt{(2\pi)^d|R|}} exp\{-\frac{1}{2} (z-Hx)^TR^{-1}(z-Hx)\}
%\label{Gau_h}
%\eq
Then $p(x,\lambda)$ is Gaussian such that
\bq
p(x,\lambda) = \frac{1}{\sqrt{(2\pi)^n|P_p|}} exp\{-\frac{1}{2} (x-x_p)^TP_p^{-1}(x-x_p)\}.
\label{Gau_p}
\eq
Furthermore, $x_p$ and $P_p$ are unique solutions to (\ref{mean1}) and (\ref{cov1}), respectively, with
$x_0=x_{prior}$ and $P_0=P_g$.
}
\vs

Note that, for the stochastic particle flow (\ref{f1})-(\ref{q1}), the matrix $Q(\lambda)$ (or $K$) does not appear in the density function $p(x,\lambda)$, the mean $\bar{x}(\lambda)$, and the covariance matrix $P(\lambda)$ at all. We will see in Section 5 that $Q(\lambda)$ plays an important role in stabilizing the flow and error reduction for applications in filtering and Bayesian inference. 

A Gaussian distribution is completely determined by its mean and covariance matrix. With the establishment Theorem 3.1, we know that the particle flow $x(\lambda)$ defined by (\ref{f1})-(\ref{q1}) has distribution $p(x,\lambda)$ for all $\lambda\in [0, 1]$, which says that the particle flows are unbiased with correct covariance matrix. Because the density function of a linear stochastic differential equation is Gaussian \cite{Arn}, Theorem 3.1 effectively establishes the equivalence, up to a free parameter $K$, of linear flow and the posterior distribution for all $\lambda \in [0, 1]$ under the Gaussian assumption.
 
With the establishment of Theorem 3.1, we know that $\hat{x}_N$ in (\ref{est_mean}) and $\hat{P}_N$ in (\ref{est_var}) are unbiased estimates of the posterior mean and covariance matrix, respectively. Furthermore, the average is taken over i.i.d. random variables in (\ref{est_mean}) and (\ref{est_var}) with finite mean and finite covariance matrix. As the number of particles $N$ goes to infinity, for given $x_{prior}$ and $z$, their convergence is guaranteed by the strong law of large numbers \cite{Shi}, i.e.,
\[ 
\lim_{N\ra\infty}\hat{x}_N = \bar{x}(\lambda)|_{\lambda=1}, \hs a.s.
\] 
and
\[ 
\lim_{N\ra\infty}\hat{P}_N = P(\lambda)|_{\lambda=1}, \hs a.s.
\]
In other words, the estimates are consistent.
\vs

Linear minimum variance estimation plays a foundational role in estimation and filtering. Kalman filtering could be recast as a recursive implementation of the linear minimum variance estimate for linear dynamic systems under Gaussian assumptions. Given two random variables $x$ and $z$, the linear minimum variance estimate is \cite{Shi}
\[
\hat{x}^* = \mu_x + R_{xz}R_{zz}^{-1}(z-\mu_z)
\]
in which $\mu_x, \mu_z$ are the mean of prior, the mean of measurement, respectively, $R_{xz}$ is the covariance matrix between $x$ and $z$, and $R_{zz}$ is the covariance matrix of $z$. When both $x$ and $z$ are Gaussian distributed, and if the measurement $z$ is linear in $x$, the linear minimum variance estimate coincides with the posterior mean of $x$ \cite{Shi}. Theorem 3.1 indicates that the mean of the particle flow (\ref{flow}) at $\lambda=1$ is the same as the linear minimum variance estimate, which is another way of establishing the consistency of the estimates (\ref{est_mean}) and (\ref{est_var}). 

It should be emphasized that the particle flow (\ref{flow}), (\ref{f1})-(\ref{q1}) defines a general form of flow for nonlinear filtering. No Gaussian assumption is required. The only requirement is the assumption (A2), which is mild. It has been observed that its performance, for nonlinear filtering problems, is superior or among the best in a wide range of applications \cite{CP,DOT,GYM,KKS,KB,MC,MCC,PC,PMZC,Port,SLC,Wu,Zhao}. The linear Gaussian case is only used for obtaining theoretical guarantees, which is important to ensuring that particle flow filters give correct answers. These results represent a major progress for particle flow filters.

\section{Special Cases}
Since the concept of particle flow filters was first introduced in \cite{DH2007}, there have been a number of particle flows proposed in the literature \cite{Daum2016}. In this section, we examine the relationship of the new parameterized family of stochastic flow (\ref{f1})-(\ref{q1}) with those existing in the literature. 
%For simplicity, we limit our discussion to the new stochastic flow (\ref{f1}) and (\ref{q2}) as a parameterized functional of $Q(\lambda)$ as long as $Q(\lambda)$ is positive semi-definite and independent of $x$.

We first introduce several particle flows that are relevant to the topic of this paper. We exclude deterministic flows but the Exact Flow because they are rather different in nature from stochastic flows. Those interested in various particle flows, deterministic or stochastic, can find them in \cite{CL,Daum2016, DHN2018}.
\vs

\noindent
{\em 1. The Exact Flow} \cite{DHN2010}. Assume that the prior $g(x)$ has probability density function (\ref{Gau_g}), the measurement is linear (\ref{lin_meas}), and $R$ is positive definite (and independent of $x$). The Exact Flow is a deterministic linear flow, for a given deterministic initial condition, and constructed as the following
\bq
\frac{dx}{d\lambda}=A_1(\lambda)x+b_1(\lambda)
\label{exactflow}
\eq
in which
\bq
A_1(\lambda) = -\frac{1}{2}P_gH^T(\lambda HP_gH^T+R)^{-1}H,
\label{ef_A}
\eq

\bq
b_1(\lambda) = (I+2\lambda A_1)[(I+\lambda A_1)P_gH^TR^{-1}z+A_1x_{prior}].
\label{ef_b}
\eq
Because an initial condition is random (a particle), the deterministic flow is actually a stochastic process without the diffusion term. The Exact Flow allows the measurement $z$ to be a nonlinear function of $x$.
\vs

\noindent
{\em 2. Stochastic Flow with Fixed $Q$} \cite{DHN2016,DHN2018}. This flow is defined by (\ref{flow}) with $f$ and $Q$ as a pair jointly chosen as the following 
\[
f = -(\nb_x\nb_x^T\log p)^{-1}(\nb_x\log h),
\]
\[
Q = -(\nb_x\nb_x^T\log p)^{-1}(\nb_x\nb_x^T\log h)(\nb_x\nb_x^T\log p)^{-1}.
\]
Note that, under the assumption that $g(x)$ and $h(x)$ are sufficiently (second order) differentiable, $\nb_x\nb_x^T\log p$ is symmetric by definition. Under the assumptions (A1) and (A2) in Theorem 2.1, $Q$ is positive semi-definite and independent of $x$.
\vs

\noindent
{\em 3. The Diagnostic Noise Flow} \cite{Daum2016}.  The Diagnostic Noise Flow does not provide an exact solution, rather an approximation. The starting point is to choose a known flow function $\hat{f}$ such that $\nb_x div(\hat{f})\approx \nb_x div(f)$ and $\nb_x\hat{f}\approx \nb_x f$. With $\hat{f}$ chosen, the Diagnostic Noise Flow is defined as
\bq
f = -(\nb_x\nb_x^T\log p)^{-1}[\nb_x\log h+\nb_x div(\hat{f})+(\nb_x\hat{f})^T(\nb_x\log p)-\beta],
\label{diag_f}
\eq
\bq
Q =\al I_{n\times n},
\label{diag_Q}
\eq
in which 
\[
\beta = \frac{\alpha}{2}\nb_x[div(\nb_x\log p)+(\nb_x\log p)^T(\nb_x\log p)],
\]
and $\al>0$ is a constant (independent of $x$) chosen using an approximation procedure such as using the least squares method \cite{CL,Daum2016}.
\vs

\noindent
{\em 4. The Approximate Flow} \cite{Daum2016}. For an arbitrary positive semi-definite matrix $Q$, the drift function is approximated as
\bq
f\approx -(\nb_x\nb_x^T\log p)^{-1}\{\nb_x\log h+\nb_x div(\hat{f})+(\nb_x\hat{f})^T(\nb_x\log p)-\nb_x[\frac{1}{2p}\nb_x^T(pQ)\nb_x]\}
\label{approx_f}
\eq
in which $\hat{f}$ is a known flow function such as the Exact Flow. This is a general flow in the sense that the matrix $Q$ could be a function of both $x$ and $\lambda$, and $\hat{f}$ could be nonlinear in $x$.
\vs

Approximating flows, the Diagnostic Noise Flow and the Approximate Flow, are intended to improve from a known flow with an analytic form. We next establish relationships between the stochastic flow (\ref{f1}) and (\ref{q2}) with the four flows listed above.
\vs

{\sc Theorem 4.1}. {\em The following relationships hold.
\bi
\item[(1).] The Exact Flow is the same as the stochastic flow (\ref{f1}) and (\ref{q2}) when $Q=0$.
% with deterministic initial condition $x|_{\lambda=0}=x_0$.
\item[(2).] The Stochastic Flow with Fixed $Q$ is the same as (\ref{f1}) and (\ref{q2}) with $K=0$.
\item[(3).] The Diagnostic Noise Flow is a special case of (\ref{f1}) and (\ref{q2}), if $\hat{f}$ and $\nb_x\log p$ are linear in $x$.
%, and
%\[
%K=\al (\nb_x\nb_x\log p)^2-(\nb_x\hat{f})^T(\nb_x\nb_x^T\log p)\in\ck.
%\]
\item[(4).] The Approximate Flow can be recast as a special case of (\ref{f1}) and (\ref{q2}) if $\hat{f}$ is linear in $x$ and $Q$ is independent of $x$.
%, and
%\[
%K=(\nb_x\nb_x^T\log p)Q(\nb_x\nb_x^T\log p) -(\nb_x\hat{f})^T(\nb_x\nb_x^T\log p)\in\ck.
%\]
\ei
}
\vs

In other words, Theorem 4.1 states that the stochastic flow  defined by (\ref{f1}) and (\ref{q1}) indeed has a general form. It unifies several flows existing in the literature. Therefore, these seemingly different flows are actually the same flow with different design choices (parameters). Both the existing Diagnostic Noise Flow and the Approximate Flow are approximate flows. 
%According to Corollary 2.1, we may choose any diagonal $Q$ as long as we modify $f$ according to (\ref{q2}). 
It should be pointed out that both the Diagnostic Noise Flow and the Approximate Flow are rather general. The Diagnostic Flow allows general $\nb_x\log p$ and $\hat{f}$ that may be nonlinear in $x$. The Approximate Flow allows for general $Q$ that could be a function of $x$, or nonlinear $\hat{f}$. However, both provide approximations of $f$ only. In this paper, we focus on exact solutions to (\ref{cond1}) and their analytical properties.

\section{Stability Analysis}
A major advantage of introducing the diffusion term in the particle flow is to stabilize particle flows (for potential error reduction) and prevent particles from diverging. 
Using noise to stabilize a system is an established approach in control systems and other applications \cite{ACW,Kha2012,Mao,Sha}. Applications to filtering problems in engineering such as target tracking have demonstrated that stochastic particle flows with nonzero diffusion significantly increase the stability of filtering performance and reduce error. Despite abundant empirical evidences there are few results on theoretical guarantees that introducing $Q$ indeed improves stability. Part of the issue is that the linearized form of the stochastic differential equation that governs the particle flows is $\lambda$-varying, which invalidates approaches to stability analysis based on eigenvalue assignments. %However, the impact of $K$ or $Q$ on the eigenvalues of the linearized form of the flow equation does affect the dynamics of the flow, which may be potentially exploited to address issues such as stiffness in the differential equation of the flow \cite{DH2014,MDD}. This aspect warrants further investigation.

To illustrate the stabilizing effect of noise in stochastic systems, we consider the following examples in time \cite{Kha2012,Koz}. For any $a>0$, the deterministic differential equation
\bq  
\frac{dx}{dt} = ax, x(0)=x_0
\label{example}
\eq
is unstable and $\lim_{t\ra\infty}x(t)=\infty$ if $x_0\neq 0$. However, consider the following stochastic differential equation by adding noise to it.
\[
dx(t) = axdt+\si xdw(t)
\]
in which $\sigma$ is a constant and $w(t)$ is a one dimensional Brownian motion with $E[(dw)^2]=1$. Its solution is
\[  
x(t) = x_0e^{(a-\si^2/2)t+\si w(t)}.
\]
Then $\lim_{t\ra\infty}x(t)= 0, a.s.$ for any $x_0$
if $a<\si^2/2$. The stochastic differential equation is stable. In other word, adding noise stabilizes an unstable deterministic system (\ref{example}) provided that the noise is sufficiently strong.
\vs

Another issue is time span. Traditionally, stability is concerned with the limiting behavior of a dynamical system near an equilibrium as time goes to infinity. Such stability concept does not directly apply to particle flows in a homotopy because $\lambda$ is limited to $[0, 1]$. Therefore, we consider finite time stability. Furthermore, the stochastic differential equation (\ref{flow}) governed by (\ref{f1})-(\ref{q1}) is $\lambda$-varying and does not take a stationary equilibrium unless for special cases (e.g., $x_{prior}=0$, $z=0$, $Q=0$). From a practical point of view, we are mostly concerned with numerical stability in implementing particle flows for nonlinear filtering or other inference problems for decision-making. Therefore, we focus on the stability of the error system.

Assume that $x_1(\lambda)$ and $x_2(\lambda)$ are two different solutions to (\ref{flow}) starting with different initial conditions. Under the assumption (A1) in Theorem 2.1, the error $\tilde{x}(\lambda)=x_1(\lambda)-x_2(\lambda)$ satisfies
\bq
d\tilde{x} = A(\lambda)\tilde{x}d\lambda, \hs \tilde{x}(0)=\tilde{x}_{0}.
\label{error}
\eq
where $\tilde{x}_0=x_1(0)-x_2(0)$ and
\bq
A(\lambda) = (\nb_x\nb_x^T\log p)^{-1}[-\nb_x\nb_x^T \log h+K].
\label{error_A}
\eq
We next examine the stability of (\ref{error}) which is in fact a deterministic system for given initial condition $\tilde{x}_0$. In practice, the initial condition $\tilde{x}_0$ represents error in estimation or prediction, thus a random variable. It is necessary to consider the stochastic aspect of the stability also.

In (\ref{error}), the coefficient matrix $A(\lambda)$ is $\lambda$-varying and related to the parameter matrix $K$ through (\ref{error_A}). For a given $\lambda \in [0, 1]$, the eigenvalues of $A(\lambda)$ may potentially be changed by choosing appropriate $K$ for each fixed $\lambda$. However, it's well established that negative eigenvalues may not guarantee stability of a time-varying system \cite{Kha}. Lyapunov-like approaches are often adopted.
\vs

%We are mostly interested in the possible values of $\tilde{x}(\lambda)$ at $\lambda=1$. Consequently, similar to the work by Kushner in \cite{Kus}, we will derive bounds for
%\[ 
%P(||\tilde{x}(\lambda)||\leq \ga),
%P(||\tilde{x}(\lambda)||\geq\ep)
%\]
%in which $\ga>0, \ep>0$ are constants, and $||.||$ is a norm. Without loss of generality, our discussion is limited to the $l_2$ norm $||.||_2$. The first quantity $P(||\tilde{x}(\lambda)||\leq \ga)$ is use for characterizing boundedness of $\tilde{x}(\lambda)$, and the second quantity $P(||\tilde{x}(\lambda)||\geq\ep)$ for characterizing how fast the error deceases to zero, both in terms of probability. Only one measure is needed because $P(||\tilde{x}(\lambda)||\leq \ga)=1.0-P(||\tilde{x}(\lambda)||< \ga)$. We keep both measures to emphasize the two different types of stability.

%The conventional concepts of finite time stability (e.g., in\cite{Dor2006}, \cite{AACC2006}) involve a finite settling time within a fixed time interval of interest. For our problem, we are concerned with flow behavior at the final time $\lambda=1$. Therefore, we modify the finite time stability concepts in \cite{AACC2006} and define concepts of finite terminal time stability.
%\vs

{\sc Definition 5.1.}\cite{AACC2006,Dor2006,Kam1953} {\em Given two positive scalars $\al, \be$, with $\al < \be$, and a symmetric positive definite constant matrix $S\in \mathbb{R}^{n\times n}$ (independent of both $x$ and $\lambda$), the system (\ref{error})
is said to be finite time stable  with respect to $(\al, \be, S)$, if
\bq 
\tilde{x}_0^T S\tilde{x}_0<\al \Rightarrow \tilde{x}^T(\lambda) S \tilde{x}(\lambda) < \be, \forall \lambda\in [0, 1].
\label{fts1}
\eq
}
\vs

{\sc Definition 5.2.}\cite{WI1965} {\em Given three positive scalars $\al, \be, \ga$, with $\al < \ga, \be<\al$, and a symmetric positive definite matrix $S \in \mathbb{R}^{n\times n}$ (independent of both $x$ and $\lambda$), the system (\ref{error})
is said to be finite time contractively stable  with respect to $(\al, \be, \ga, S)$, if it is finite time stable with respect to $(\al,\ga,S)$ and there exists a $\lambda_1\in (0,1)$ such that
\bq 
\tilde{x}^T(\lambda) S\tilde{x}(\lambda) < \be, \forall \lambda\in [\lambda_1, 1].
\label{fts2}
\eq
}

The concepts of finite time stability and finite time contractive stability are analogous (finite time version) to conventional bounded stability and asymptotic stability, respectively. However, conventional Lyapunov stabilities and the finite time stabilities are in general independent concepts \cite{AACC2006}.
\vs

{\sc Definition 5.3.}\cite{Kus} {\em Given three positive scalars $\al, \be, \ep, 0<\ep<1$, and a symmetric positive definite matrix function $S\in \mathbb{R}^{n\times n}$, the system (\ref{error}) is said to be finite time stochastically stable  with respect to $(\al, \be, \ep, S)$, if 
\bq 
E[\tilde{x}_0^T S \tilde{x}_0] \leq \al \Rightarrow \Pro[\tilde{x}^T(\lambda) S \tilde{x}(\lambda)\leq\be, \forall \lambda\in [0, 1]]\geq 1-\ep.
\label{fts3}
\eq
}

In practice, we prefer both $\beta$ and $\ep$ in (\ref{fts3}) as small as possible.
\vs

%Generally speaking, Definitions 5.1-5.3 are similar to Lyapunov stability concepts but they are independent concepts. A finite time stability does not imply certain Lyapunov stability, and vice versa \cite{AACC2006}.
%\vs

{\sc Lemma 5.1}. {\em Assume the assumptions (A1) and (A2) in Theorem 2.1. Let $M(\lambda)=-\nb_x\nb_x^T\log p$ and define a continuous function 
\bq
V(\lambda,\tilde{x}) = \tilde{x}^TM\tilde{x} \in R^+.
\label{V}
\eq
Then for the stochastic particle flow defined by (\ref{flow}), (\ref{f1}), (\ref{q1}), we have
\bq
dV(\lambda,\tilde{x})= - (M\tilde{x})^TQ(M\tilde{x})d\lambda.
\label{dV}
\eq
}
%\hspace{0.1in}\rule{1.5mm}{3.0mm}

Under the assumptions (B1) and (B2) in Theorem 5.1 below, $V(\lambda,\tilde{x})$ defines a Lyapunov function \cite{Kal}. However, there usually does not exist a constant $\dt>0$ such that $V(\lambda,\tilde{x})\leq \dt||\tilde{x}||$ for all $\lambda \geq 0$ if $\lambda$ is allowed to go to infinity because $M$ increases linearly in $\lambda>0$. 
\vs

{\sc Theorem 5.1.} {\em Assume that 
\bi
\item[(B1).] $\nb_x\log g$ and $\nb_x\log h$ are linear in $x$,
\item[(B2).] $\nb_x\nb_x^T\log g$ is negative definite and $\nb_x\nb_x^T\log h$ is negative semi-definite.
\ei
Then there exists a symmetric positive definite matrix $S\in \mathbb{R}^{n\times n}$ such that the error system (\ref{error}) is finite time stable with respect to $(\al,\beta,S)$ for any $\al>0, \beta>0, \al<\beta$. 
}
\vs

{\sc Theorem 5.2.} {\em Assume the assumptions (B1) and (B2) in Theorem 5.1 and
\bi
\item[(B3).] $Q(\lambda)\geq Q_0$ for all $\lambda\in[0, 1]$ and $Q_0$ is a constant positive definite matrix, 
\ei
then there exists a symmetric positive definite matrix $S\in \mathbb{R}^{n\times n}$ and a positive scalar $\si>0$ such that the error system (\ref{error}) is finite time contractively stable with respect to $(\al,\beta,\ga,S)$ for all positive scalars $\al, \beta, \ga$ satisfying $\al<\ga, \beta<\al$ as long as $\al e^{-\si}<\beta<\al$.
}
\vs

{\sc Theorem 5.3.} {\em Under the assumptions (B1)-(B3), there exists a symmetric positive matrix $S\in \mathbb{R}^{n\times n}$ such that for any positive scalars $\al>0, \beta>0, \ep>0$,
(\ref{error}) is finite time stochastically stable with respect to $(\al, \beta, \ep, S)$ as long as $\al<\beta, \al/\beta \leq\ep<1$.
}
\vs

Theorems 5.1-5.3 are general stability results. The equation (\ref{dV}) reveals details about the dynamic behaviors of particle flows. Denote $S=-\nb_x\nb_x^T\log g\in \mathbb{R}^{n\times n}$. Then $S$ is symmetric positive definite under the assumption (B2). Note that $\nb_x\nb_x^T\log p = \nb_x\nb_x^T\log g+\lambda\nb_x\nb_x^T\log h$. We may write
\[
M(\lambda) = S+\lambda (-\nb_x\nb_x^T\log h), \forall \lambda \in [0, 1]
\]
which shows that $M$ increases linearly in $\lambda\in\mathbb{R}^+$ since $-\nb_x\nb_x^T\log h$ is positive semi-definite under the assumption (B2). For the Exact Flow, $Q(\lambda)=0$. The follow conclusion follows directly from (\ref{dV}).
\vs

{\sc Theorem 5.4}. {\em Assume the assumption (B2) in Theorem 5.1. For the Exact Flow, we have
\[
\tilde{x}(\lambda)^TM(\lambda)\tilde{x}(\lambda) = \tilde{x}_0^T S\tilde{x}_0, \forall \lambda \in [0, 1],
\]
in other words, if we start with $N$ particles, $x_i(0), i=1, 2, ...,N$, and if the errors for all particles start on an ellipsoid $E_0\df\{x \in \mathbb{R}^n | x^TSx=\al, \al>0\}$, they will remain on the same but smaller ellipsoid $E_\lambda\df\{x \in \mathbb{R}^n | x^TM(\lambda)x=\al, \al>0\}$ for all $\lambda\in [0, 1]$.}
\vs

It is sometimes useful to have a Lyapunov-like quantity to characterize the behavior of $\tilde{x}$. Under the assumptions (B1)-(B3) in Theorems 5.1-5.2, we know from (\ref{a_V3}) in the Appendix that
\[
\tilde{x}^T(\lambda) S \tilde{x}(\lambda) \leq 
e^{-\si\lambda}\tilde{x}_0^T S \tilde{x}_0, \forall \lambda \in [0, 1].
\]
where $S=-\nb_x\nb_x^T\log g>0, \si=\lambda_{min}(Q_0)\lambda_{min}(S)>0$. Note that $\tilde{x}_0$ represents error in initial condition. This equation states that the error contracts exponentially at least at the rate of $\si$. In assumption (B3), $Q_0$ is the lower bound of $Q$. Increasing the minimum eigenvalue of $Q_0$ increases the guaranteed speed at which the error system (\ref{error}) goes to zero.
\vs

%We use $M(\lambda)$ in defining Lyapunov function. 
Note that $M(\lambda)\geq S$. We know that, 
\[
  \tilde{x}^TM(\lambda)\tilde{x}\leq 0 \Rightarrow \tilde{x}^TS\tilde{x}\leq 0.
\]
Table 1 summarizes the relationship between $Q$ and error dynamics due to initial conditions.
%, which indicates that choosing appropriate $Q$ may potentially improve filtering accuracy by reducing the effects of initial error $\tilde{x}_0$.

\begin{table}[h!]
	\renewcommand{\arraystretch}{1.3}
	\begin{center}
		\caption{Relationship between $Q(\lambda)$ values and error dynamics}
		\vs
		\label{tab:table1}
		\begin{tabular}{|c|c|c|}
			\hline
			 & Lyapunov function  & Error dynamics over $\lambda\in[0, 1]$\\ 
			\hline
			$Q(\lambda)=0$ & $\tilde{x}^TM(\lambda)\tilde{x}$ &   $=$ constant \\
			\hline
			$Q(\lambda)=0$ & $\tilde{x}^TS\tilde{x}$, $E[\tilde{x}^TS\tilde{x}]$ &   non-increasing \\
			\hline
			$Q(\lambda)\geq 0$ & $\tilde{x}^TM(\lambda)\tilde{x}$, $\tilde{x}^T S\tilde{x}$, $E[\tilde{x}^TS\tilde{x}]$ & non-increasing \\
			\hline
			$Q(\lambda)\geq Q_0>0$ & $\tilde{x}^T S\tilde{x}$, $E[\tilde{x}^TS\tilde{x}]$ & exponentially decreasing at rate $\geq\sigma>0$ \\
			\hline
		\end{tabular}
	\end{center}
\end{table}

Table 1 shows that the flow is “more" stable for $Q\geq 0$ than that for $Q=0$, which explains observations that have been reported in the implementations of particle flow filters in several important applications \cite{CL,DH2007,DHN2016,DHN2018,GYM}. Furthermore, a strictly positive definite $Q>0$ is generally desirable to minimize the propagation/accumulation of errors in initial conditions for the flow. In this sense, our results in this section provide guidelines to the design of particle flow filters to ensure robust performance.

\section{Conclusions}
In this paper, we have addressed two aspects of stochastic particle flows: (1) derivation of a new parameterized family of stochastic flows driven by a nonzero diffusion process. It is shown that several existing stochastic flows and the deterministic Exact Flow are special cases of this family. This family includes all possible stochastic flow of linear forms driven by constant noise that is independent of state. (2) initial work toward laying a theoretical foundation for the particle flow filters. In particular, we established unbiasedness of the particle flows with correct covariance matrix, the consistency of estimates derived from stochastic particle flows, its connection with linear minimum variance estimation theory, and numerical stability in implementation.

The contributions of this paper are (1) theoretical proofs that particle flow filters give correct answers, (2) a parameterized form of particle flow that unifies seemingly disconnected particle flows in the literature as special choices of design parameters, and (3) guidelines on the stability of particle flows to ensure robust performance for nonlinear filtering. The contributions are significant in themselves and also in that for the first time we demonstrated the feasibility toward establishing a comprehensive theoretical foundation for particle flow filters. There are multiple possibilities for future efforts. A comprehensive list of topics for future research is provided in \cite{DH2015,DH2016}. Here we just point out a few directly relevant to the topic of this paper. In this paper we focus on stochastic flows. The results should be extended to deterministic flows that are derived using other methods such as Gromov's method. Details about deterministic flows can be found in \cite{DH2015,DHN2018}. %Another possibility is to relax the assumption that $Q$ is independent of state. Choosing $Q$ as a general function of state has played an important role in the stabilization of stochastic systems in fields such as control systems. Efforts in this direction could expand the applicability of particle flows to broader applications such as controller design for nonlinear systems. 
In Section 5, we demonstrated the role of $Q$ in stabilizing the error system (\ref{error}) and in error reduction. The parameter matrix $K$ or $Q$ may be exploited to change the dynamics of the flow equation (\ref{flow}) to address issues such as stiffness of (stochastic/ordinary) differential equations toward improving numerical stability of filtering performance. Stiffness of differential equations is an important issue in, for example, certain tracking problems \cite{DH2014}. The linearized form of the stochastic differential equation (\ref{flow}) is (\ref{lin1}) in which the eigenvalues of the coefficient matrix $A(\lambda)$ can be changed by choosing appropriate $K$ or $Q$, which in turn affects the stiffness of the flow. Further research is warranted to characterize the exact relationship between the choice of $K$ and the stiffness of the flow. % Lastly, to achieve theoretical traceability, linear Gaussian assumptions are adopted as in the theoretical studies of Kalman filters and particle filters. It'd be beneficial to relax such assumptions in the analysis of nonlinear filtering.

\section{Acknowledgment}
LD would like to thank Timothy Zajic for inspiring discussions regarding finding all possible solutions to (\ref{cond1}).

\newpage
\begin{center}
	{\Large \bf Appendix}\\[2mm]
\end{center}

We first state several lemmas that will be needed in the proofs.
\vs
 
\noindent
{\sc Lemma A.1.} Let $A(\theta)\in \mathbb{R}^{n\times n}$ be a parameterized non-singular square matrix and differentiable with respect to the parameter $\theta\in \mathbb{R}$ in a neighborhood of a given point. Then
\[   \frac{dA^{-1}(\theta)}{d\theta} = - A^{-1}(\theta)(\frac{dA(\theta)}{d\theta})A^{-1}(\theta).
	\]
%\bi
%\item[(a.1)]
%\[   \frac{dA^{-1}(\theta)}{d\theta} = - A^{-1}(\theta)(\frac{dA(\theta)}{d\theta})A^{-1}(\theta)
%	\]
%\item[(a.2)]
%\[
%\frac{d|A(\theta)|}{d\theta} = |A(\theta)|tr(A^{-1}(\theta)\frac{dA(\theta)}{d\theta})
%\]
%\ei
\vs

\noindent
{\sc Lemma A.2.}\cite{Ho, Jaz} Assume all consistent dimensions and existence of matrix inversions. The following two matrix equalities hold.
%p.119.
%Matrices in control theory:?
\bi
\item[(a.1)] A matrix equality
\[
%(A+BCD)^{-1}BC = A^{-1}B(C^{-1}+DA^{-1}B)^{-1}.
(A^{-1}+M^TB^{-1}M)^{-1}M^TB^{-1}=AM^T(MAM^T+B)^{-1}.
\]
\item[(a.2)] Woodbury's Matrix Inversion Lemma
\[
%(A+BCD)^{-1}=A^{-1}-A^{-1}B(C^{-1}+DA^{-1}B)^{-1}DA^{-1}.
(A^{-1}+M^TB^{-1}M)^{-1}=A-AM^T(MAM^T+B)^{-1}MA.
\]
\ei
\vs

\noindent
{\sc Lemma A.3.}\cite{Bel,Gron} (Gronwall-Bellman Inequality) 
Let $u(t)\in \mathbb{R}$ be a differentiable, positive function and $\al(t)\in \mathbb{R}$ is a continuous function. If the derivative of $u(t)$ satisfies
\[
\frac{d u(t)}{dt}\leq \al(t)u(t), \forall t\geq t_0
\]
then
\[  u(t)\leq u(t_0)e^{\int_{t_0}^t\al(s)ds}, \forall t \geq t_0.
\]
In particular, if $\al(t)=-\al$, $\al>0$ a constant,
\[u(t)\leq u(t_0)e^{-\al(t-t_0)}, \forall t \geq t_0.
\]
\vs

\noindent
{\sc Lemma A.4.} \cite{Shi} (Markov Inequality) 
Let $x\in \mathbb{R}^+$ be a nonnegative random variable and $\al>0$. Then
\[
E[x\geq \al]\leq \frac{E[x]}{\al}.
\]
\vs

%\noindent
%{\sc Proof of Lemma 2.1}
%
%The condition (\ref{cond1}) was derived in the literature, see for example \cite{DH2013}. We sketch the derivation for self-containedness. 
%We start with taking the partial derivative with respect to $\lambda$ on both sides of (\ref{loghom}) to obtain
%\bq
% \log h - \frac{d\log c(\lambda)}{d\lambda} = \frac{1}{p}\frac{\partial p}{\partial \lambda}.
%\label{a_dloghom}
%\eq
%On the other hand, for the stochastic process (\ref{flow}), the Kolmogorov's forward equation gives \cite{Jaz}
%\[
%\frac{\partial p(x,\lambda)}{\partial \lambda} =
%-div(pf)+\frac{1}{2}\sum_{i,j=1}^n\frac{\partial^2 [p(Q_{i,j})]}{\partial x_i\partial x_j}
%= -div(pf)+\frac{1}{2}\nb_x^T(pQ)\nb_x.
%\]
%Substituting the previous expression into the right hand side of (\ref{a_dloghom}), we have
%\[
%\log h - \frac{d\log c(\lambda)}{d\lambda}
%= -\frac{1}{p}div(pf)+\frac{1}{2p}\nb_x^T(pQ)\nb_x.
%=-div(f)-(\nb_x \log p)^Tf+\frac{1}{2p}\nb_x^T(pQ)\nb_x.
%\]
%Taking gradient with respect to $x$ on both sides of the previous equation gives (\ref{cond1}). We thus complete the proof of Lemma 2.1.
%\vs

\noindent
{\sc Proof of Theorem 2.1} 

Under the assumption (A1), $\nb_x\nb_x^T\log p$ is not a function of $x$.
If $Q$ is not a function of $x$ (could be a function of $\lambda$), it is shown in \cite{CL} that
\bq
  \nb_x[\frac{1}{2p}\nb_x^T(pQ)\nb_x] =(\nb_x\nb_x^T\log p)Q(\nb_x\log p).
\label{a_eqQ}
\eq
We next consider $f$ as a linear combination of $\nb_x\log p$ and $\nb_x\log h$:
\bq
    f = K_1\nb_x\log p +K_2\nb_x\log h.
\label{a_f}
\eq
Under the assumption (A1), $f$ is linear in $x$, thus
\bq \nb_x div(f)=0.
\label{a_div}
\eq 
Substituting (\ref{a_eqQ})-(\ref{a_div}) into (\ref{cond1}), we get
\[ \nb_x\log h=-(\nb_x\nb_x^T\log p)[K_1\nb_x\log p+K_2\nb_x\log h] \]
\bq
-[(\nb_x\nb_x^T\log p)K_1^T+(\nb_x\nb_x^T\log h) K_2^T](\nb_x\log p)+(\nb_x\nb_x^T\log p)Q(\nb_x\log p).
\label{a_cond2}
\eq
By setting the coefficient matrices corresponding to $\nb_x\log p$ and $\nb_x\log h$ equal on both sides, we obtain
\bq
I = -(\nb_x\nb_x^T\log p)K_2,
\label{a_eq_K1}
\eq

\bq
0 = -(\nb_x\nb_x^T\log p)K_1
-(\nb_x\nb_x^T\log p)K_1^T-(\nb_x\nb_x^T\log h) K_2^T+(\nb_x\nb_x^T\log p)Q.
\label{a_eq_K2}
\eq
Under the assumption (A2), $\nb_x\nb_x^T\log p$ is invertible. It follows from (\ref{a_eq_K1}) that
\bq  
K_2 = - (\nb_x\nb_x^T\log p)^{-1}.
\label{a_K2}
\eq
Substituting this form of $K_2$ into (\ref{a_eq_K2}), we have
\[
(\nb_x\nb_x^T\log p)Q=
(\nb_x\nb_x^T\log p)(K_1+K_1^T)
-(\nb_x\nb_x^T\log h) (\nb_x\nb_x^T\log p)^{-1}
\]
or
\bq
Q= K_1+K_1^T
-(\nb_x\nb_x^T\log p)^{-1}(\nb_x\nb_x^T\log h) (\nb_x\nb_x^T\log p)^{-1}.
\label{a_Q}
\eq
Setting
\[ 
K = (\nb_x\nb_x^T\log p)^{-1}K_1 (\nb_x\nb_x^T\log p)^{-1},
\]
then we know from (\ref{a_f}), (\ref{a_K2}), and (\ref{a_Q}) that such choices of $K$ and $Q$ satisfy (\ref{f1}) and (\ref{q1}). This completes the proof of Theorem 2.1.
\vs

\noindent
{\sc Proof of Corollary 2.1}

We only consider symmetric matrix $K$. Then it follows from (\ref{q1}) that
\[
Q = (\nb_x\nb_x^T\log p)^{-1}(-\nb_x\nb_x^T\log h+2K)(\nb_x\nb_x^T\log p)^{-1}.
\]
Solving for $K$, we get
\[ 
K= \frac{1}{2}(\nb_x\nb_x^T\log p)Q(\nb_x\nb_x^T\log p)+\frac{1}{2}\nb_x\nb_x^T\log h
\]
which is exactly (\ref{q2}).
\vs

\noindent
{\em Proof of Theorem 2.2}

Under the assumption (A1) in Theorem 2.1, $f\in\cf$ is linear in $x$. Assume that $\hat{f}$ is another linear function of $x$ that also solves (\ref{cond1}). Then
\bq
\nb_x\log h =-(\nb_x\nb_x^T\log p)\hat{f} - \nb_x div(\hat{f})-(\nb_x \hat{f})^T(\nb_x\log p)+\nb_x[\frac{1}{2p}\nb_x^T(pQ)\nb_x].
\label{a_cond1}
\eq 
Let $\tilde{f}=\hat{f}-f$. Subtracting (\ref{a_cond1}) from (\ref{cond1}), and also noticing $\nb_xdiv(\tilde{f})=0$ since $\tilde{f}$ is linear in $x$, we obtain
\bq
0 =-(\nb_x\nb_x^T\log p)\tilde{f} -(\nb_x \tilde{f})^T(\nb_x\log p).
\label{a_eqf2}
\eq
Note that $\tilde{f}$ is linear in $x$, we may write
\[
\tilde{f}=\tilde{A}(\lambda)x+\tilde{b}(\lambda).
\]
Substituting this form into (\ref{a_eqf2}), we get
\bq
0=-(\nb_x\nb_x^T\log p)\tilde{f} -\tilde{A}^T(\lambda)(\nb_x\log p).
\label{a_quad}
\eq
or
\[
0=-(\nb_x\nb_x^T\log p)[\tilde{A}(\lambda)x+\tilde{b}(\lambda)] -\tilde{A}^T(\lambda)(\nb_x\log p).
\]
The right hand side of the previous equation is linear in $x$. The coefficient matrix of $x$ must be zero in order to have zero on its left hand side, which gives
\bq
(\nb_x\nb_x^T\log p)\tilde{A}(\lambda) +\tilde{A}^T(\lambda)(\nb_x\nb_x^T\log p)=0.
\label{a_Atilde}
\eq
Under the assumption (A2), $\nb_x\nb_x^T\log p$ is invertible. We solve for $\tilde{f}$ from (\ref{a_quad}) to obtain,
\[ 
\tilde{f} = -(\nb_x\nb_x^T\log p)^{-1}\tilde{A}^T(\lambda)(\nb_x\log p).
\]
Its combination with the form of $f$ in (\ref{f1}) leads to
\bq
  \hat{f}=f+\tilde{f} 
=(\nb_x\nb_x^T\log p)^{-1}[-\nb_x\log h+K'(\nb_x\nb_x^T\log p)^{-1}(\nb_x\log p)]
\label{a_ftilde}
\eq
in which 
\[
 K'= K-\tilde{A}^T(\lambda)(\nb_x\nb_x^T\log p).
\]
For $K\in \ck$, we know from (\ref{a_Atilde}) that
\[
K'+(K')^T-\nb_x\nb_x^T\log h
\]
\[
=K+K^T-(\nb_x\nb_x^T\log p)\tilde{A}(\lambda) -\tilde{A}^T(\lambda)(\nb_x\nb_x^T\log p)-\nb_x\nb_x^T\log h
\]
\[
=K+K^T-\nb_x\nb_x^T\log h\geq 0,
\] 
or $K'\in\ck$. Therefore, $\hat{f}\in\cf$ according to (\ref{a_ftilde}). This completes the proof of Theorem 2.2.
\vs

\noindent
{\sc Proof of Theorem 3.1}

It should be pointed out that by construction, the probability density function of the particle flow $x$ in (\ref{flow}) is $p(x,\lambda)$. In the following, we verify that it is indeed the case. 
%We verify that the probability density function of the particle flow $x$ in (\ref{flow}) indeed has the posterior distribution $p(x,\lambda)$ as specified in (\ref{homotopy}) for all $\lambda\in[0, 1]$.
%No proof is really needed. But the verification provides an additional layer of confidence for the results.

For the linear measurement (\ref{lin_meas}) with $v$ being Gaussian distributed, we have
\bq
h(x) = \frac{1}{\sqrt{(2\pi)^d|R|}} exp\{-\frac{1}{2} (z-Hx)^TR^{-1}(z-Hx)\}.
\label{a_Gau_h}
\eq
According to (\ref{homotopy}), we have
\[
p(x,\lambda) = \frac{g(x)h^{\lambda}(x)}{c(\lambda)}
\]
\bq =\frac{1}{c_1(\lambda)}\exp\{-\frac{1}{2}(x-x_{prior})^TP_g^{-1}(x-x_{prior})-\frac{\lambda}{2}(z-Hx)^TR^{-1}(z-Hx)\}
\label{a_Gau_p1}
\eq
where $c_1(\lambda)$ is the normalization term (a constant). Note that the exponent in (\ref{a_Gau_p1}) is quadratic in $x$.
It is straightforward to verify that we can rewrite the exponent as
\[
-\frac{1}{2}(x-x_{prior})^TP_g^{-1}(x-x_{prior})-\frac{\lambda}{2}(z-Hx)^TR^{-1}(z-Hx)
\]
\[
=-\frac{1}{2}x^T(P_g^{-1}+\lambda H^TR^{-1}H)x+x^T(P_g^{-1}x_{prior}+\lambda H^TR^{-1}z)+\textrm{constant terms}
\]
\[ =-\frac{1}{2}(x-x_\mu)^TP_p^{-1}(x-x_\mu)-\log c(\lambda)
\]
in which $c(\lambda)$ is a constant (normalization) term,
\bq
x_\mu(\lambda) = (P_g^{-1}+\lambda H^TR^{-1}H)^{-1}(P_g^{-1}x_{prior}+\lambda H^TR^{-1}z)
\label{a_p_mean}
\eq
and
\bq
P_p(\lambda) = (P_g^{-1}+\lambda H^TR^{-1}H)^{-1}.
\label{a_p_var}
\eq
We next verify that $x_\mu(\lambda)$ and $P_p(\lambda)$ satisfy (\ref{mean1}) and (\ref{cov1}), respectively, for any $K$.

According to (\ref{Gau_g}) and (\ref{a_Gau_h}), we have
\[
\log g = -\frac{1}{2}(x-x_{prior})^TP_g^{-1}(x-x_{prior})-\log\sqrt{(2\pi)^n|P_g|}
\]
and
\[
\log h = -\frac{1}{2}(z-Hx)^TR^{-1}(z-Hx)-\log\sqrt{(2\pi)^d|R|}.
\]
Therefore,
\bq
\nb_x\log g = -P_g^{-1}(x-x_{prior}), \hs \nb_x\nb_x^T\log g = - P_g^{-1}
\label{a_logh}
\eq
and
\bq
\nb_x\log h = H^TR^{-1}(z-Hx), \hs \nb_x\nb_x^T\log h = -H^TR^{-1}H.
\label{a_logg}
\eq
Using the relationship (\ref{loghom}),
\bq
\nb_x\log p = \nb_x\log g+\lambda\nb_x\log h=-P_g^{-1}(x-x_{prior})+\lambda H^TR^{-1}(z-Hx),
\label{a_logp}
\eq
\bq
\nb_x\nb_x^T\log p = \nb_x\nb_x^T\log g+\lambda\nb_x\nb_x^T\log h = -P_g^{-1}-\lambda H^TR^{-1}H.
\label{a_d2logp}
\eq
According to (\ref{a_p_mean}) and (\ref{a_p_var}), we can re-write $x_{\mu}(\lambda)$ as
\[ 
x_{\mu}(\lambda) = P_p(P_g^{-1}x_{prior}+\lambda H^TR^{-1}z).
\]
Setting $\bar{x}=x_{\mu}(\lambda)$,  the left hand side (LHS) of (\ref{mean1}) is
\[
\textrm{LHS of (\ref{mean1})} = \frac{dx_{\mu}(\lambda)}{d\lambda} 
\]
\[
= (\frac{dP_p}{d\lambda})(P_g^{-1}x_{prior}+\lambda H^TR^{-1}z)+P_p H^TR^{-1}z
\]
Applying the differentiation of matrix inverse formula in Lemma A.1, also noticing the form of $P_p$ in (\ref{a_p_var}), we have
\[ 
\textrm{LHS of (\ref{mean1})} = [-P_p(H^TR^{-1}H)P_p](P_g^{-1}x_{prior}+\lambda H^TR^{-1}z)+P_pH^TR^{-1}z
\]
\[
= -P_p(H^TR^{-1}H)x_{\mu}+P_pH^TR^{-1}z
\]
\[
= P_p[H^TR^{-1}(z-Hx_{\mu})]
\]
\bq
= (\nb_x\nb_x^T\log p)^{-1}(-\nb_x\log h)|_{x=x_{\mu}}.
\label{a_LHS_mean}
\eq
On the other hand, note that
\[
\nb_x\log p |_{x=x_{\mu}} \equiv 0, \forall \lambda \in [0, 1].
\]
For the right hand side (RHS) of (\ref{mean1}), we know
\[
\textrm{RHS of (\ref{mean1})} = f_{|x=x_{\mu}}
\]
\[
=(\nb_x\nb_x^T\log p)^{-1}[-\nb_x\log h+K(\nb_x\nb_x^T\log p)^{-1}(\nb_x\log p)]|_{x=x_{\mu}}
\]
\[
=(\nb_x\nb_x^T\log p)^{-1}(-\nb_x\log h)|_{x=x_{\mu}} 
\]
\[= \textrm{(LHS) of (\ref{mean1})},
\]
according to (\ref{a_LHS_mean}). As for the initial condition, according to (\ref{a_p_mean}), we have
\[
x_{\mu}(\lambda)|_{\lambda=0}=x_{prior}.
\]
We have thus established (\ref{mean1}).

Next we examine both sides of (\ref{cov1}). For $P_p(\lambda)$ defined in (\ref{a_p_var}), using the formula in Lemma A.1,
\bq
\textrm{LHS of (\ref{cov1})} = \frac{d P_p(\lambda)}{d\lambda}
= - P_p H^TR^{-1}H P_p
\label{a_LHS_P}
\eq
According to (\ref{lin1_A}),
\bq
A(\lambda) = (\nb_x\nb_x^T\log p)^{-1}[-\nb_x\nb_x^T\log h+K] 
%=- (P_g^{-1}+\lambda H^TR^{-1}H)^{-1}[H^TR^{-1}H+K]
= -P_p(H^TR^{-1}H+K)
\label{a_A}
\eq
%\[
%= -P_p(H^TR^{-1}H+K)
%\]
and consequently, 
\[
\textrm{RHS of (\ref{cov1})} = -P_p(H^TR^{-1}H+K)P_p-P_p(H^TR^{-1}H+K^T)P_p+Q
\]
\[
=-P_p(H^TR^{-1}H+K)P_p-P_p(H^TR^{-1}H+K^T)P_p+
P_p(H^TR^{-1}H+K+K^T)P_p
\]
\[
= -P_pH^TR^{-1}HP_p = \textrm{LHS of (\ref{cov1})}.
\]
As for the initial condition, we have
\[ P_p|_{\lambda=0} = P_g.
\]
which establishes (\ref{cov1}). 
Since $p(x,\lambda)$ is Gaussian whose density function is completely determined by its mean and covariance matrix, we thus complete the proof of Theorem 3.1.
%According to (\ref{lin1_b}),
%\[
%b(\lambda)=f-A(\lambda)x=(\nb_x\nb_x^T\log p)^{-1}[-H^TR^{-1}z+K (\nb_x\nb_x^T\log p)^{-1}(P_g^{-1}x_{prior}+\lambda H^TR^{-1}z)]
%\]
\vs

\noindent
{\em Proof of Theorem 4.1}

\noindent
(1). The Exact Flow was obtained in \cite{DHN2010} by directly solving (\ref{cond1}) with $Q=0$ and under linear Gaussian assumptions. It was derived again in \cite{CL,Khan}. We need to verify that when $Q=0$, (\ref{lin1_A}) and (\ref{lin1_b}) become (\ref{ef_A}) and (\ref{ef_b}), respectively.
When $Q=0$, we know from (\ref{q2}) that
\[ 
K = \frac{1}{2}(\nb_x\nb_x^T\log h).
\]
Substituting this form into (\ref{lin1_A}), we get
\[
A(\lambda) = (\nb_x\nb_x^T\log p)^{-1}[-\nb_x\nb_x^T \log h+\frac{1}{2}\nb_x\nb_x^T \log h]
\]
\bq
= -\frac{1}{2}(\nb_x\nb_x^T\log p)^{-1}\nb_x\nb_x^T \log h.
\label{a_A2}
\eq
Under Gaussian assumptions, we know from (\ref{a_logh}) and (\ref{a_d2logp}) that
\bq
A(\lambda) =-\frac{1}{2}(P_g^{-1}+\lambda H^TR^{-1}H)^{-1}H^TR^{-1}H.
\label{a_A3}
\eq
Applying the matrix equality in Lemma A.2 (a.1), with $A=P_g, M= H, B=\lambda^{-1}R$, we can rewrite 
\[ 
(P_g^{-1}+\lambda H^TR^{-1}H)^{-1}H^TR^{-1} 
= P_gH^T(\lambda HP_gH^T+R)^{-1}.
\]
Substituting this matrix identity into the form of $A(\lambda)$ in (\ref{a_A3}), we obtain
\[
A(\lambda) = -\frac{1}{2}[P_gH^T(\lambda HP_gH^T+R)^{-1}]H
=A_1(\lambda).
\]
%in which the last equality follows from the matrix inversion equation in Lemma A.2 (a.1).
As for $b(\lambda)$ in (\ref{lin1_b}),
\[ 
b(\lambda) = f-A(\lambda)x
\]
\[
=(\nb_x\nb_x^T\log p)^{-1}[-\nb_x\log h+K(\nb_x\nb_x^T\log p)^{-1}(\nb_x\log p)]-A(\lambda)x
\]
\[
=(\nb_x\nb_x^T\log p)^{-1}[-\nb_x\log h+\frac{1}{2}(\nb_x\nb_x^T\log h)(\nb_x\nb_x^T\log p)^{-1}(\nb_x\log p)]-A(\lambda)x.
\]
Applying the forms of $\nb_x\log h$ in (\ref{a_logh}) and $\nb_x\log g$ in (\ref{a_logg}), also noting the form of $A(\lambda)$ in (\ref{a_A2}), we have
\[
b(\lambda)=(\nb_x\nb_x^T\log p)^{-1}[-H^TR^{-1}z +\frac{1}{2}(\nb_x\nb_x^T\log h)(\nb_x\nb_x^T\log p)^{-1}(P_g^{-1}x_{prior}+\lambda H^TR^{-1}z)]
\]
\[
=(\nb_x\nb_x^T\log p)^{-1}(-H^TR^{-1}z)-A(\nb_x\nb_x^T\log p)^{-1}(P_g^{-1}x_{prior}+\lambda H^TR^{-1}z)]
\]
\bq
=(I+\lambda A)(P_g^{-1}+\lambda H^TR^{-1}H)^{-1}H^TR^{-1}z + A(P_g^{-1}+\lambda H^TR^{-1}H)^{-1}P_g^{-1}x_{prior}.
\label{a_b1}
\eq
By applying the Woodbury's matrix inversion lemma, with $A=P_g, M= H, B=\lambda^{-1}R$ in Lemma A.2 (a.2), we have
\[ (P_g^{-1}+\lambda H^TR^{-1}H)^{-1}
= P_g-\lambda P_gH^T(R+\lambda HP_gH^T)^{-1}HP_g
\]
\[
= (I+2\lambda A_1)P_g = (I+2\lambda A)P_g.
\]
Substituting it into (\ref{a_b1}), we obtain
\[
b(\lambda) = (I+\lambda A)(I+2\lambda A)P_gH^TR^{-1}z+A(I+2\lambda A)x_{prior}
\]
\[
= (I+2\lambda A)[(I+\lambda A)P_gH^TR^{-1}z+Ax_{prior}]= b_1(\lambda).
\]
We have thus proved that the Exact Flow is the same as (\ref{f1}) and (\ref{q2}) when $Q=0$ (this is the only case of (\ref{flow}) being deterministic).
\vs

\noindent
(2). This is a special case of (\ref{f1}) and (\ref{q2}) since this particular $Q$ leads to $K=0$.
\vs

\noindent
(3). For $\nb_x\log p$ linear in $x$, we have
\[
\nb_x[div(\nb_x\log p)] = 0, \hs \nb_x[(\nb_x\log p)^T(\nb_x\log p)]=2(\nb_x\nb_x^T\log p)(\nb_x\log p)
\]
in which $\nb_x\nb_x^T\log p$ is independent of $x$. Consequently,
\[
\beta =\al (\nb_x\nb_x^T\log p)(\nb_x\log p).
\]
If $\hat{f}$ is linear in $x$, $\nb_x div(\hat{f})=0$ and $\nb_x\hat{f}$ is independent of $x$. Therefore, noting that $Q=\al I$, the $f$ in (\ref{diag_Q}) becomes
\[ f = - (\nb_x\nb_x^T\log p)^{-1}[\nb_x\log h+(\nb_x\hat{f})^T(\nb_x\log p)-\al (\nb_x\nb_x^T\log p)(\nb_x\log p)]
\]
\[
= (\nb_x\nb_x^T\log p)^{-1}[-\nb_x\log h+K(\nb_x\nb_x^T\log p)^{-1}(\nb_x\log p)]\in\cf
\]
where 
\bq
K=[(\nb_x\nb_x^T\log p)Q -(\nb_x\hat{f})^T](\nb_x\nb_x^T\log p).
\label{a_K_1}
\eq
To complete the proof, we need to verify that $K\in\ck$. 
Since $\hat{f}$ is a linear particle flow, it satisfies (\ref{cond1}). Therefore,
\[
\nb_x\log h=-(\nb_x\nb_x^T\log p)\hat{f}-(\nb_x\hat{f})^T(\nb_x\log p)
+(\nb_x\nb_x^T\log p)Q(\nb_x\log p).
\]
Taking gradient on both sides of the previous equation, we get
\[
\nb_x\nb_x^T\log h=-(\nb_x\nb_x^T\log p)(\nb_x\hat{f})-(\nb_x\hat{f})^T(\nb_x\nb_x^T\log p)
+(\nb_x\nb_x^T\log p)Q(\nb_x\nb_x^T\log p).
\]
or
\bq
\nb_x\nb_x^T\log h+(\nb_x\nb_x^T\log p)(\nb_x\hat{f})+(\nb_x\hat{f})^T(\nb_x\nb_x^T\log p)
=(\nb_x\nb_x^T\log p)Q(\nb_x\nb_x^T\log p).
\label{a_K_2}
\eq
According to (\ref{a_K_1}),
\[ 
-\nb_x\nb_x^T\log h+K+K^T
\]
\[
=-\nb_x\nb_x^T\log h+[(\nb_x\nb_x^T\log p)Q -(\nb_x\hat{f})^T](\nb_x\nb_x^T\log p)+(\nb_x\nb_x^T\log p)[Q\nb_x\nb_x^T\log p-\nb_x\hat{f}]
\]
\bq
=-[\nb_x\nb_x^T\log h +(\nb_x\hat{f})^T(\nb_x\nb_x^T\log p)+(\nb_x\nb_x^T\log p)\nb_x\hat{f}]+2
(\nb_x\nb_x^T\log p)Q(\nb_x\nb_x^T\log p).
\label{a_K_3}
\eq
Combining (\ref{a_K_3}) and (\ref{a_K_2}) gives \[-\nb_x\nb_x^T\log h+K+K^T=(\nb_x\nb_x^T\log p)Q(\nb_x\nb_x^T\log p)> 0,
\]
or $K\in\ck$. Thus this is a special case of (\ref{f1}) and (\ref{q2}). %Note that the Diagnostic Noise Flow is an approximation and satisfies (\ref{cond1}) if $K\in\ck$.
\vs

\noindent
(4). If $\hat{f}$ is linear in $x$, $\nb_xdiv(\hat{f}$)=0 and $\nb_x\hat{f}$ is independent of $x$. Furthermore, for $Q$ independent of $x$, we know from (\ref{a_eqQ}) that
\[ 
\nb_x[\frac{1}{2p}\nb_x^T(pQ)\nb_x]=(\nb_x\nb_x^T\log p)Q(\nb_x\log p).
\]  
Consequently, (\ref{approx_f}) becomes
\[
f\approx -(\nb_x\nb_x^T\log p)^{-1}\{\nb_x\log h+[(\nb_x\hat{f})^T-(\nb_x\nb_x^T\log p)Q](\nb_x\log p)\}.
\]
\[
=(\nb_x\nb_x^T\log p)^{-1}[-\nb_x\log h+K(\nb_x\nb_x^T\log p)^{-1}(\nb_x\log p)]\in\cf
\]
in which 
$K=[(\nb_x\nb_x^T\log p)Q -(\nb_x\hat{f})^T](\nb_x\nb_x^T\log p)\in\ck$ as proved in the previous case of the Diagnostic Noise Flow. Thus this is a special case of (\ref{f1}) and (\ref{q2}).

This completes the proof of Theorem 4.1.
\vs

\noindent
{\em Proof of Lemma 5.1}

According to (\ref{loghom}), $M(\lambda) = -\nb_x\nb_x^T\log g -\lambda \nb_x\nb_x^T\log h$.
Then $dM/d\lambda=-\nb_x\nb_x^T\log  h$. Therefore,
\[   dV(\lambda,\tilde{x}) 
%= \tilde{x}^TA^T(\lambda)M(\lambda)\tilde{x}d\lambda+\tilde{x}^T(dM/d\lambda)\tilde{x}d\lambda+\tilde{x}^TM(\lambda)A(\lambda)\tilde{x}d\lambda
%\]
%\[
= \tilde{x}^T[A^T(\lambda)M(\lambda)-\nb_x\nb_x^T\log h+M(\lambda)A(\lambda)]\tilde{x}d\lambda.
\]
The coefficient matrix $A(\lambda)$ is defined in (\ref{error_A}), and can be written as $A(\lambda)=-M^{-1}(\lambda)[-\nb_x\nb_x^T\log h+K]$. Consequently
\[ 
dV(\lambda,\tilde{x}) =-\tilde{x}^T[-\nb_x\nb_x^T\log h+K+K^T]\tilde{x}d\lambda.
\]
According to (\ref{q1}), 
\[
Q=(\nb_x\nb_x^T\log p)^{-1}(-\nb_x\nb_x^T\log h+K+K^T)(\nb_x\nb_x^T\log p)^{-1}
\]
or equivalently,
\[
-\nb_x\nb_x^T\log h+K+K^T = (\nb_x\nb_x^T\log p)Q(\nb_x\nb_x^T\log p).
\]
Therefore,
\[
dV(\lambda,\tilde{x})
=-\tilde{x}^T(\nb_x\nb_x^T\log p)Q(\nb_x\nb_x^T\log p)\tilde{x}d\lambda
\]
\[ =-\tilde{x}^TMQM\tilde{x}d\lambda=-(M\tilde{x})^TQ(M\tilde{x})d\lambda
\]
which is exactly (\ref{dV}).
\vs

\noindent
{\em Proof of Theorem 5.1}

Let $S=-\nb_x\nb_x^T\log g\in\mathbb{R}^{n\times n}$. Then $S$ is a constant symmetric matrix. Under the assumption (B2), $S$ is positive definite.
The matrix $Q$ is positive semi-definite. Lemma 5.1 shows that $dV/d\lambda\leq 0$ for any given $\tilde{x}_0$, or
\bq
V(\lambda,\tilde{x}(\lambda))\leq V(0,\tilde{x}(0))=\tilde{x}_0 S\tilde{x}_0, \forall \lambda \in [0, 1]. 
\label{a_V1}
\eq
On the other hand, under the assumption (B2), $-\nb_x\nb_x^T\log h$ is positive semi-definite. For any $\lambda\in [0, 1]$,
\bq 
V(\lambda,\tilde{x})=\tilde{x}^TS\tilde{x}+\lambda\tilde{x}^T(-\nb_x\nb_x^T\log h)\tilde{x}
\geq\tilde{x}^T S \tilde{x}.
\label{a_V2}
\eq
Combining (\ref{a_V2}) with (\ref{a_V1}), we know that
\bq
\tilde{x}^T S\tilde{x}\leq \tilde{x}_0^T S \tilde{x}_0, \forall \lambda \in [0, 1].
\label{a_V22}
\eq
With this $S$ chosen, for any $\al>0, \beta>0, \al<\beta$, $\tilde{x}_0^T S\tilde{x}_0<\al$ leads to
\[ 
\tilde{x}^T S\tilde{x} \leq \tilde{x}_0^T S \tilde{x}_0< \al<\beta, \forall \lambda \in [0, 1],
\]
in other words, the error system (\ref{error}) is finite time stable with respect to $(\al,\beta,S)$.
\vs

\noindent
{\em Proof of Theorem 5.2}

Let $S=-\nb_x\nb_x^T\log g\in\mathbb{R}^{n\times n}$. The matrix $S$ is constant, symmetric, positive definite under assumption (B2). The matrix $M=-\nb_x\nb_x^T\log p=S+\lambda(-\nb_x\nb_x^T\log h)$ is positive definite for all $\lambda\in [0, 1]$ since $S$ is positive definite and $-\nb_x\nb_x^T\log h$ is positive semi-definite according to assumption (B2). System (\ref{error}) is finite time stable with respect to $(\al,\ga,S)$ according to Theorem 5.1. Furthermore, under the assumptions (B2)-(B3), we have
\[  
\tilde{x}^TMQM\tilde{x}
\]
\[
\geq \tilde{x}^TMQ_0 M\tilde{x}
\textrm{ \hs (according to assumption (B3) $Q\geq Q_0$)}
\]
\[\geq \lambda_{min}(Q_0)\tilde{x}^TM M\tilde{x} \textrm{ \hs ($\lambda_{min}(Q_0)$ is the minimum eigenvalue of $Q_0$)}
\]
\[
= \lambda_{min}(Q_0)\tilde{y}^TM\tilde{y} 
\textrm{ \hs (setting $\tilde{y} = M^{1/2} \tilde{x}$, $M$ is positive definite by assumption (B2))}
\]
\[=\lambda_{min}(Q_0)\tilde{y}^T[S+\lambda(-\nb_x\nb_x^T\log h)] \tilde{y}
\textrm{ \hs (from definiton of $M$)}
\]
\[
\geq \lambda_{min}(Q_0)\tilde{y}^T S\tilde{y}
\textrm{ \hs ($-\nb_x\nb_x^T\log h$ is positive semi-definite and $\lambda\geq 0$)}
\]
\[\geq \lambda_{min}(Q_0)\lambda_{min}(S)\tilde{y}^T\tilde{y}
\textrm{ \hs ($\lambda_{min}(S)$ is the minimum eigenvalue of $S$)}
\]
\[
=\lambda_{min}(Q_0)\lambda_{min}(S)\tilde{x}^TM\tilde{x}
\textrm{ \hs (return to $\tilde{x}$ and $M$ is symmetric)}
\]
\[
=\lambda_{min}(Q_0)\lambda_{min}(S)V(\lambda,\tilde{x})
\textrm{ \hs (definition of $V$)}
\]
\[
=\si V(\lambda,\tilde{x})
\]
in which $\si=\lambda_{min}(Q_0)\lambda_{min}(S)>0$ under the assumptions (B2)-(B3). Combining the above inequality with (\ref{dV}) in Lemma 5.1, we know that 
\bq
\frac{dV}{d\lambda} \leq - \tilde{x}^TMQM\tilde{x} \leq -\si V, \forall \lambda \in [0, 1].
\label{a_dv}
\eq
According to Lemma A.3, (\ref{a_dv}) gives
\[ 
V(\lambda,\tilde{x})\leq e^{-\si\lambda}V(0,\tilde{x}_0)=e^{-\si\lambda}\tilde{x}_0^T S\tilde{x}_0, \hs \forall \lambda \in [0, 1].
\]
Combining this inequality with (\ref{a_V2}), we get
\bq
\tilde{x}^T(\lambda) S \tilde{x}(\lambda) \leq 
e^{-\si\lambda}\tilde{x}_0^T S\tilde{x}_0, \hs \forall \lambda \in [0, 1].
\label{a_V3}
\eq
For any $\al>0, \beta>0$ satisfying $\al e^{-\si}<\beta<\al$, we can select a $\lambda_1$, $0<\lambda_1<1$, sufficiently close to $1$ such that $\al e^{-\si\lambda_1}<\beta$. With this $\lambda_1$ chosen, for any $\tilde{x}_0$ satisfying $\tilde{x}_0^TS\tilde{x}_0<\al$,
(\ref{a_V3}) guarantees that
\[\tilde{x}^T(\lambda)S\tilde{x}(\lambda) < 
e^{-\si\lambda_1}\al<\beta, \hs \forall \lambda \in [\lambda_1, 1]
\]
which establishes that (\ref{error}) is finite time contractively stable as stated in Theorem 5.2.
\vs

\noindent
{\em Proof of Theorem 5.3}

Let $S=-\nb_x\nb_x^T\log g\in\mathbb{R}^{n\times n}$. Then $S$ is positive definite under assumption (B2) and $\tilde{x}^TS\tilde{x}>0$ for all $\tilde{x}\neq0$, $=0$ if $\tilde{x}=0$. Therefore,  $\tilde{x}^TS\tilde{x}$ is a nonnegative random variable and
\[  
Prob[\tilde{x}^T S\tilde{x}\geq\beta]
\]
\[\leq
\frac{E[\tilde{x}^T S\tilde{x}]}{\beta}
\textrm{ \hs (applying Markov Inequality in Lemma A.4)}
\]
\[
\leq \frac{E[\tilde{x}_0^T S\tilde{x}_0]}{\beta}
\textrm{ \hs (according to (\ref{a_V22}))}
\]
\[ \leq \al/\beta
\]
for all $\tilde{x}_0$ satisfying $E[\tilde{x}_0^T S\tilde{x}_0]\leq \al$ and $\lambda \in [0, 1]$.
Consequently, 
\[
Prob[\tilde{x}^TS\tilde{x}\leq\beta]
\geq 1-\al/\beta \geq 1-\ep\]
as long as $\al<\beta$ and $0<\ep<=\al/\beta$.
We thus complete the proof of Theorem 5.3.
\vs

\noindent
{\em Proof of Theorem 5.4}

Theorem 5.4 follows directly from (\ref{dV}) for $Q=0$. 

\end{document}